# MULTI-SCALE CHARACTERISATION OF MONUMENT LIMESTONES

K. Beck[1,2], O. Rozenbaum[1*], M. Al-Mukhtar[2], A. Plançon[1]

[1]Institut des Sciences de la Terre d'Orléans, Université d'Orléans, CNRS-ISTO, 1A rue de la Férollerie, 45071 Orléans Cedex 2, France.
[2]Centre de Recherche sur la Matière Divisée, Université d'Orléans, CNRS-CRMD, 1B rue de la Férollerie, 45071 Orléans Cedex 2, France.
*corresponding author: e-mail: rozenbaum@cnrs-orleans.fr

**ABSTRACT**
*Among the parameters influencing stone deterioration, moisture and water movements through the pore network are essential. This communication presents differents methods to characterize stones and to determinate the water transfer properties. Results are analysed for two limestones having similar total porosity, but characterized by different pore networks. These different porous systems govern dissimilar water properties.*

## INTRODUCTION

Water plays a fundamental role in the phenomena of stone deterioration. The understanding of water movements within the pore network of stone, which affect the behaviour of porous materials, are essential. The study of porous media and water transfer in fresh stones is a first step in order to analyse and to model deterioration process. In this paper, two sedimentary limestones are chosen. The first one, called tuffeau, is a stone commonly used in most of monuments built along the Loire valley in France. Meanwhile, it is necessary to note the great diversity of the tuffeau family (important variability of minerals proportion and porosity [1]). So, studied tuffeau is tuffeau of Saumur, extracted from an underground quarry, and the other one, Sébastopol stone, is obtained from a quarry near Paris.

## CHARACTERIZATION OF THE STUDIED STONES

### THE SOLID PHASE
The mineralogical composition of studied stones is obtained by several complementary techniques : X-ray diffraction, infrared and Raman spectroscopies, thermogravimetric analysis, ICP, EMPA coupled with analysis from optical microscopy.

In tuffeau and Sébastopol stone as in many sedimentary rocks, the main crystalline phases are calcite $CaCO_3$ and silica $SiO_2$ in the form of quartz. But other minerals are found in the tuffeau as clays (e.g. glauconie which appears greenish in optical microscopy) and micas visible to the naked eye, and especially another crystalline form of silica named opal cristobalite-tridymite (opal CT). It is also interesting to note the presence of a small quantity of minerals which appear with a metal luster as seen in optical microscopy. Indeed, very small grains of titanium oxyde $TiO_2$ are found rather uniformly spread, both in the tuffeau and the Sébastopol stone. Other detritic minerals like zircon are also found in small grains spread in the Sébastopol stone.

Calcium carbonate is distributed among sparitic and micritic calcite, and fossil shells. The calcite proportion is quantified using the mass loss resulting from calcite decomposition around 700°C. The weight ratio of $CaCO_3$ is 49.6% for tuffeau and 79.5% for Sébastopol stone. So, one can classify Sébastopol stone as a limestone while the tuffeau should be classified as a siliceous limestones [2].

The grain morphology is determined from SEM images on fragments of stones (Figs. 1 and 2). The difference in the grain sizes for the two stones is clearly observed. The grains of the Sébastopol stone, as well as the marine fossils being there, are rather homogeneous and large, while size and shape of the constituent grains of the tuffeau are varied but finer. Indeed, due to the multi-phase aspect, tuffeau is constituted with grains of very different sizes and shapes. One can find in the tuffeau rather voluminous grains of quartz, but also aggregates of small crystals of micritic calcite, small fossil shells, and objects formed during the sedimentation: the opal CT spherules with their characteristic surface [3]. The texture of stones observed by SEM images allows to see the complexity of the porous network generated with these grains.

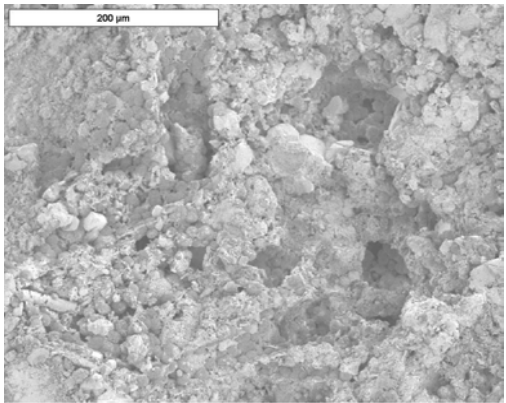 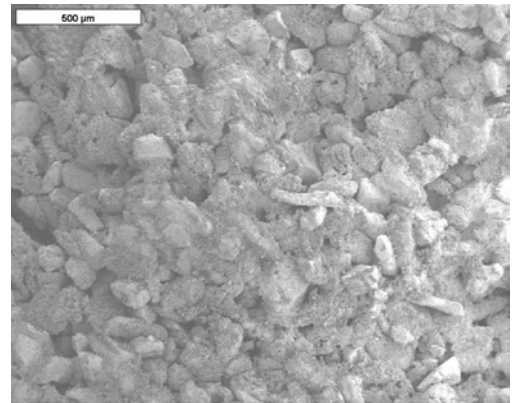

Fig. 1. SEM image of tuffeau        Fig. 2. SEM image of Sébastopol stone

## THE POROUS PHASE

The pore space is investigated by various methods like mercury porosimetry, BET measurements and 2D images analysis.

The porosity is determined from dry bulk and skeletal densities measured by hydrostatic weighing (table 1). Measurements by Helium pycnometry confirm these results ($\rho_s$ = 2.53 g/cm$^3$ for tuffeau and $\rho_s$ = 2.70 g/cm$^3$ for Sébastopol stone). Values of total porosity are near 45% for both stones, which indicates that these stones are highly porous and so particularly light building materials.

|  | Bulk dry density $\rho_b$ | Skeletal density $\rho_s$ | Porosity |
|---|---|---|---|
| Tuffeau | 1.307 | 2.545 | 48.6% |
| Sébastopol stone | 1.576 | 2.713 | 41.9% |

Table 1 : densities and porosity

Mercury porosimetry give information about pore sizes distribution (Figs. 3 and 4). For both stones, results are dissimilar: tuffeau has a very wide range of pore sizes (from 20 µm to 7 nm), which demonstrate clearly the multi-scale nature of tuffeau pore space relative to Sébastopol stone [2], whereas the major part of pore sizes for Sébastopol stone is restricted from 40 μm to 1 μm.

The total porosity are nearly the same, but due to differences between mineralogical constitution, grain sizes and shape arrangements, their pore space structure are completely different. Sébastopol stone is constitued with large grains, so the pore space is large. Tuffeau is constitued by various small grains with various morphologies like clays, micas, opal CT spherules. The arrangement of larger grains generate macro-porosity, and micro-porosity is formed by the finest phases' morphology and arrangement.

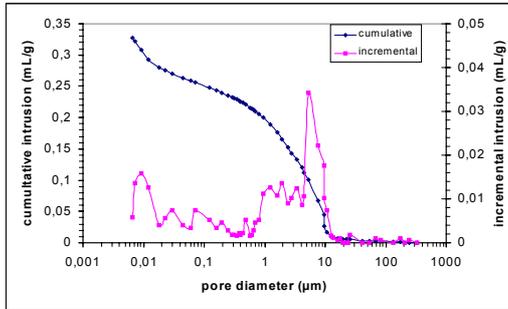 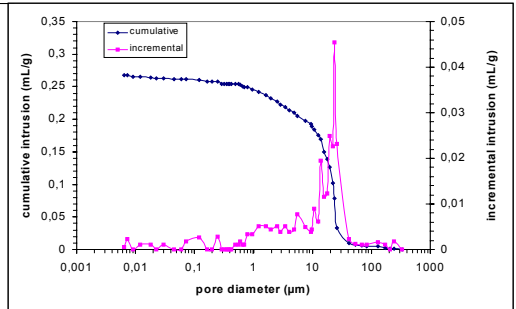

Fig. 3. Pore size distribution of tuffeau from mercury porosimetry    Fig. 4. Pore size distribution of Sébastopol stone from mercury porosimetry

The backscattered SEM images of polished cuttings allow to show the various kinds of pores (Figs. 5 and 6). These images give additional information to those obtained by the mercury porosimetry technique: for Sébastopol stone, the size of major pores is near 20 μm whereas the SEM photograph show several pores with a diameter near 100 μm. This is mainly due to the shape hypothesis and the model of cylindrical pores in mercury porosimetry analysis. And, there is also the "ink-bottle" effect which reduces the volume of the large pores to the benefit of small pores [4]. The term diameter in mercury porosimetry is then very relative as it characterizes only the entry diameter to the pore access. This effect is more visible for Sébastopol stone because the large pores seem be connected by narrowings proportionately smaller compared to tuffeau.

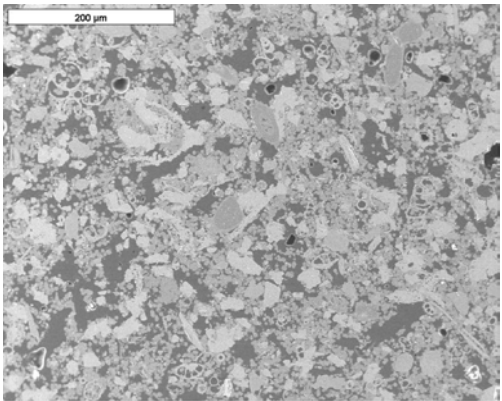 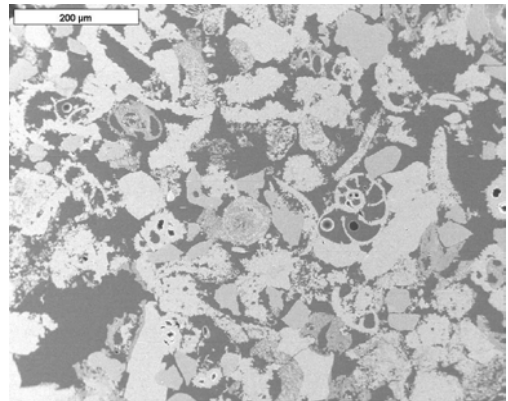

Fig. 5. Backscattered SEM photograph of tuffeau (porosity : 48.6%)    Fig. 6. Backscattered SEM photograph of Sébastopol stone (porosity : 41.9%)

These classical methods will be completed with 3D images obtained by X-ray tomography. Indeed, this tool allows to get informations about the connectivity and the topology of the pore network. Furthermore, 3D information will be useful for the modeling of water transfer in the porous media. Finally, NMR is also informative to investigate the liquid water diffusion coefficient and consequently determine the tortuosity of pore network.

## WATER TRANSFER PROPERTIES

The water retention measurement gives information about the ability of the stone to trap water from its atmospheric environment (Figs. 7 and 8). In this experiment, samples previously dried at 105°C were placed in atmospheres of different constant relative humidities kept by saturated salt solutions.

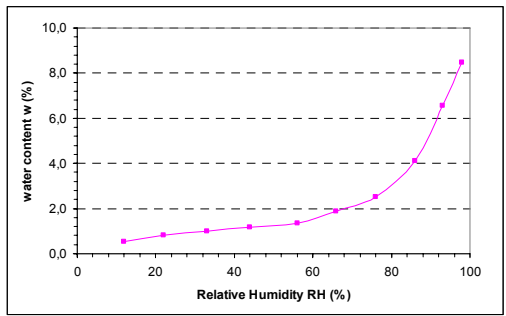 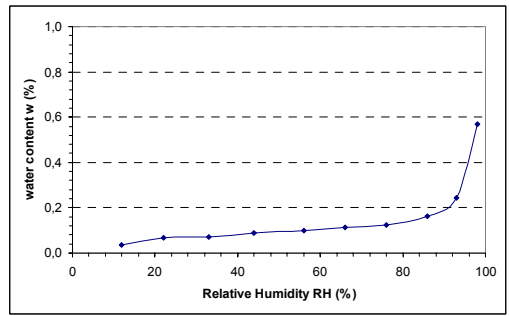

Fig. 7. Water retention curve of tuffeau     Fig. 8. Water retention curve of Sébastopol stone

The tuffeau traps easily the humidity from air and can then contain a lot of water mainly for the highest humidities. This is due to the important micro-porosity and to the presence of clayey minerals. In comparaison, the Sébastopol stone adsorbs very few amount of water, and even in very wet atmospheres, the stone remains practicaly dry. This behaviour is due to the absence of clay minerals and to a pore distribution almost exclusively constituted with very large pores.
These water retention measurements can be completed up to saturation by means of others techniques like osmotic solutions (polyetylene glycol) and tensiometric plates, imposing smaller capillary pressures [5].

The macroscopic water movement, essential for deterioration study, is mainly due to liquid water transfer. And this water is absorbed by capillary imbibition. The phenomenon of capillarity is directly related to the pore network characteristics: pore sizes and pore shapes, but also the connectivity and the topology of the porous structure.
Theoretical capillary model [6] is based on absorption by capillarity for a vertical cylindrical tube of limited height (allowing to neglect gravity). The Washburn's equations give the capillary height (h) and the mass uptake (dm) per surface unit (S) during an elapsed time (t):

$$h = \sqrt{\frac{r\sigma}{2\eta}}\sqrt{t} = B\sqrt{t}$$

$$\frac{dm}{S} = A\sqrt{t}$$

where η is the kinematic water viscosity and σ is its superficial tension.

In this test, samples, in contact with water, are weighed and the height of the capillary front are measured according to the elapsed time. In spite of the cylindrical capillary approximation, the curves of capillary ascent and mass uptake drawn according to the square root of the elapsed time (not shown here) are linear. This indicates that the pore network is homogeneous [7].
These stones have relatively high capillary kinetics (table 2) compared to less porous stones with smaller pore diameter [8], but the imbibition coefficients of the Sébastopol stone are about 2 times higher than those of the tuffeau. This confirms the influence of pore size on the imbibition kinetics. These results should be correlated to SEM images, the Sébastopol stone possesses larger pores relative to those of the tuffeau.

Imbibition leads also to determine the anisotropy of these porous media (table 2). The imbibition kinetics being faster in the direction parallel to the bedding direction. The Sébastopol stone presents a capillary anisotropy of about 20 %, while the tuffeau presents an anisotropy of about 15 %.

|  | Mass uptake coefficient A ($g/cm^2/min^{1/2}$) | Capillary front coefficient B ($cm/min^{1/2}$) |
|---|---|---|
| Tuffeau | ⊥ stone bed : 0.36<br>// stone bed : 0.42 | ⊥ stone bed : 0.96<br>// stone bed : 1.13 |
| Sébastopol stone | ⊥ stone bed : 0.62<br>// stone bed : 0.77 | ⊥ stone bed : 2.20<br>// stone bed : 2.72 |

Table 2 : imbibition coefficients in relation to bedding directions

This liquid water transfer study will be completed with vapour transfer analysis by determination of evaporation kinetics and diffusion experiments. Measurements of water vapour diffusion are carried out according the water content at steady state. This allows to calculate vapour diffusion coefficient and hydraulic conductivity variations as a function of the degree of saturation of stones [9].

## FINAL REMARKS
In spite of similar total porosities, these two stones possess very different pore size distributions. This induces different behaviours relative to water. Indeed, tuffeau and Sébastopol stone are very porous limestones, with a total porosity near 45%. But, the morphology of the solid phase and the porous phase are very dissimilar. Most of the grains of Sébastopol stone are large, whereas grains of tuffeau are smaller and varied with different size and shape (sparitic and micritic calcite, clay, mica, opal CT spherules). The arrangement of the grains in Sébastopol stone is mainly restricted to macropores whereas tuffeau has a very wide range of pore sizes with a relatively important micro-porosity. This difference is observed in the water retention curve and imbibition properties. The tuffeau traps easily humidity from air whereas Sébastopol stone is absolutely not hygroscopic. However imbibition kinetics of Sébastopol stone are about 2 times higher.
To understand stone deterioration, it is necessary at least to study water transfer properties of the stones and water effect on them. And, to understand its water transfer behaviour, it is necessary to understand first the geometry of its porous phase.

# BIBLIOGRAPHIC REFERENCES